\documentclass[aps,pra,twocolumn,reprint,amsmath,amssymb,graphicx, superscriptaddress,groupedaddress]{revtex4}

\newcommand{\beq}{\begin{equation}}
\newcommand{\eeq}{\end{equation}}
\newcommand{\bea}{\begin{eqnarray}}
\newcommand{\eea}{\end{eqnarray}}
\providecommand{\abs}[1]{\left\lvert#1\right \rvert}

\providecommand{\bra}[1]{\langle #1 \rvert}
\providecommand{\ket}[1]{\lvert #1 \rangle}

\providecommand{\tr}[1]{\text{Tr}\left[ #1 \right]}

\providecommand{\sP}{\mathcal{P}}
\providecommand{\sQ}{\mathcal{Q}}
\providecommand{\LH}{L_{\tilde{H}}}

\newcommand{\ketbra}[2]{\left| {#1} \right\rangle\left\langle {#2}\right|}

\newcommand{\un}{\openone}

\newcommand{\rank}[1]{\text{rank}\left[ #1 \right]}
\renewcommand{\ker}[1]{\text{ker} \left[ #1 \right]}
\renewcommand{\dim}[1]{\text{dim} \left[ #1 \right]}

\usepackage{amsthm}

\newtheorem{theorem}{Theorem}

\usepackage{tikz}
\usepgflibrary{arrows}
\usetikzlibrary{decorations}
\usetikzlibrary{decorations.pathmorphing,patterns}
\usetikzlibrary{scopes}
\usetikzlibrary{shapes, matrix}
\begin{document}

\title{Classification of Dark States in Multi-level  Dissipative Systems}

\author{Daniel Finkelstein-Shapiro}
\affiliation{Division of Chemical Physics, Lund University, Box 124, 221 00 Lund, Sweden}
\email{daniel.finkelstein_shapiro@chemphys.lu.se}
\author{Simone Felicetti}
\affiliation{Laboratoire Mat\'eriaux et Ph\'enom\`enes Quantiques,
Universit\'e Paris Diderot, CNRS UMR 7162, 75013, Paris, France.}
\author{Thorsten Hansen}
\affiliation{Department of Chemistry, University of Copenhagen, DK 2100 Copenhagen, Denmark}
\author{T\~onu Pullerits}
\affiliation{Division of Chemical Physics, Lund University, Box 124, 221 00 Lund, Sweden}
\author{Arne Keller}
\affiliation{Laboratoire Mat\'eriaux et Ph\'enom\`enes Quantiques,
Universit\'e Paris Diderot, CNRS UMR 7162, 75013, Paris, France.
Universit\'e Paris-Sud, 91405 Orsay, France}
\email{arne.keller@u-psud.fr}

\begin{abstract}
Dark states are eigenstates or steady-states of a system that are decoupled from the radiation. Their use, along with associated techniques such as Stimulated Raman Adiabatic Passage, has extended from atomic physics where it is an essential cooling mechanism, to more recent versions in condensed phase where it can increase the coherence times of qubits.
These states are often discussed in the context of unitary evolution and found with elegant methods exploiting symmetries, or via the Bruce-Shore transformation. However, the link with dissipative systems is not always transparent, and distinctions between classes of CPT are not always clear. We present a detailed overview of the arguments to find stationary dark states in dissipative systems, and examine their dependence on the Hamiltonian parameters, their multiplicity and purity. We find a class of dark states that depends not only on the detunings of the lasers but also on their relative intensities. We illustrate the criteria with the more complex physical system of the hyperfine transitions of $^{87}$Rb and show how a knowledge of the dark state manifold can inform the preparation of pure states. 
\end{abstract}

\maketitle

\section{Introduction}

Coherent population trapping (CPT) and transfer in few-level systems consists of the preparation of pure states or the coherent transfer among them by the use of control fields and auxiliary excited levels \cite{Radmore1982,Bergmann1998, Vitanov2017}. This arrangement is imune to certain damping processes.
The main mechanisms, CPT and Stimulated Raman Adiabatic Passage (STIRAP) have been most studied in the $\Lambda$ three-level system consisting of two-ground states and one excited state where the ground-excited transitions can be independently addressed.
There are two viewpoints of CPT depending on whether one includes dissipation or not. In a Hamiltonian system with unitary evolution, asymmetric linear combinations of the ground states decouple from a radiation field provideds some conditions are met for the field frequencies. In a dissipative system, this asymmetric combination becomes the stationary state regardless of initial conditions.
Originating in atomic and molecular physics \cite{Vanier1998,Sevincli2011}, it can be used for atomic cooling, metrology \cite{Vanier2003,Vanier2005,Guerandel2007}, testing QED, and it has been extended to state initialization in quantum information \cite{Dantan2006,Schempp2010} including qubit gates, and solid-state systems such as spin systems in nitrogen vacancies \cite{Santori2006}, superconducting circuits \cite{Kelly2010}, semiconductor heterostructures \cite{Michaelis2006,Xu2008,Issler2010,Yale2013} where coherence lifetimes have been successfully extended by orders of magnitude \cite{Chow2016, Ethier-Majcher2017}. Some recent proposals suggest its use in plasmonic systems \cite{Rousseaux2016}. 
%
%
Multiple level systems are also quite common in atomic and solid-state devices \cite{Han2008, Ivanov2010,Schempp2010,Chow2016, Ethier-Majcher2017,Vitanov2017,Gu2006}.
In order to find the solutions to dark states various approaches have been followed.
An analysis based on symmetries and constants of the motion has permitted the generalization to multilevel systems as long as the system preserves some symmetries notably with respect to the decouplings  \cite{Hioe1981,Elgin1980}. The Morris-Shore transformation separates a set of $N_g$ degenerate ground states and $N_e$ degenerate excited states into pairs of bright states, dark states and spectator states \cite{Morris1983,Rangelov2006,Shore2014}, recently extended to the case of small detunings \cite{Vasilev2014}. The transformation relies on having a number of ground states in excess in order to result in dark states. There are other instances, however, where certain states might decouple from the radiation that fall outside of these considered cases (for example for equal number of ground and excited states).


In this work, we give an overview for the conditions to find dark states in dissipative systems in Lindblad form, and systematically study the multiplicity of dark stationary states, their purity and their dependence on laser field detunings and Rabi frequencies.

\section{General considerations}
\subsection{System and definitions}
\begin{figure}[h]
\includegraphics[width=0.5\textwidth]{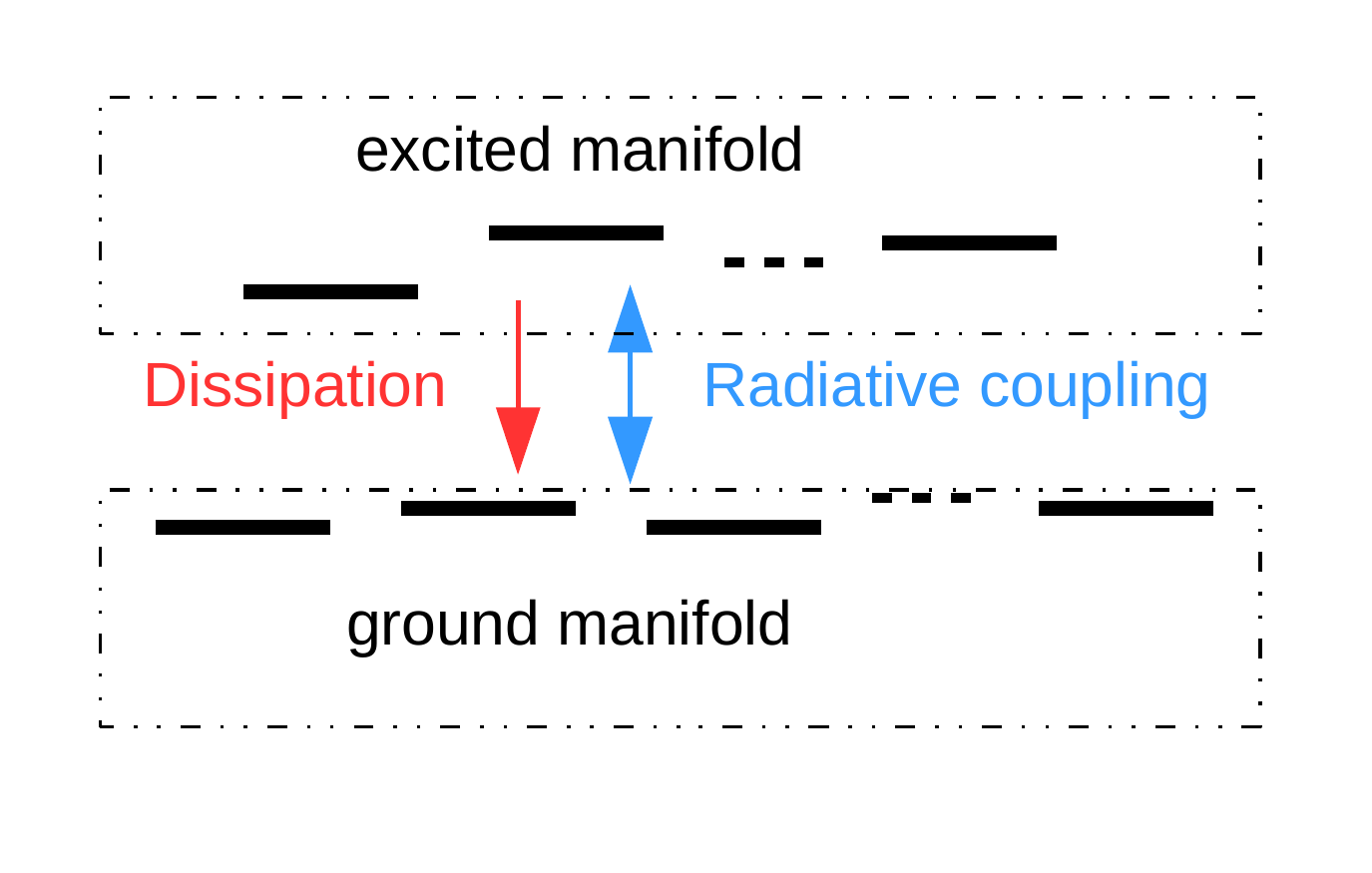}
	\caption{$N$ level system consisting of a ground state manifold coupled radiatively to an excited state manifold. The excited state can also relax dissipatively to the ground state.}
	\label{fig:manifolds}
\end{figure}

 We consider an $N$-level systems wich can be divided into $N_g$ ground states $\ket{g_i}$ with energy $E^{(g)}_i$ $(i=1,2,\cdots, N_g)$, and $N_e=N-N_g$ excited states $\ket{e_j}$ with energy $E^{(e)}_j$ $(j=1,2,\cdots, N_g)$, that decay to the ground states via spontaneous emission or coupling to a bath, at rate $\gamma_{ij}$, (see Figure \ref{fig:manifolds}).
The Hamiltonian of the system can be written as~:
\begin{align}
H(t) &= \sum_{i=1}^{N_g} E_i^{(g)} \ket{g_i}\bra{g_i} +
 \sum_{j=1}^{N_e} E_j^{(e)} \ket{e_j}\bra{e_j} + \nonumber \\
&\sum_{i,j} F_{ij}(t)\left(\ket{g_i}\bra{e_j} + \ket{e_j}\bra{g_i}\right),
\end{align}
where $F_{ij}(t) = 2F_{ij} \cos(\omega_{ij}t +\phi_{ij})$ represent the coherent laser beams with  frequencies $\omega_{ij}$ coupling the manifold  of ground states to the manifolds of excited states.



 We restrict our attention to the case where each transition $g_i\leftrightarrow e_j$ is driven by at most one laser, so that the problem can be reduced to a time-independent one (see appendix~\ref{app:RWA}). We assume no spectral overlap between the transitions.

We suppose that there exists an interaction picture such that in the rotating wave approximation (RWA),
the  system evolution is described by a Lindblad equation~\cite{Lindblad1976,Gorini1976} $\dot{\rho} = L \rho$, with a time independent Lindblad
operator $L$. The operator $L$ can be written as:
\begin{equation}
L(\rho) =-i[H,\rho]+\sum_{\{ij\}} \gamma_{ij} D_{ij}(\rho).
\label{eq:Lindblad}
\end{equation}
We set $\hbar = 1$ throughout the paper, and consider that energy and frequency are equivalent.
The Hamiltonian $H$ is now time independent  and can be written in terms of the operators $\sigma_{ij}^z = \ket{g_i}\bra{g_i} - \ket{e_j}\bra{e_j}$ (see appendix~\ref{app:RWA}) as:
\beq
\label{eq:HRWA}
H = \sum_{i=1}^{N_g}\sum_{j=1}^{N_e} \left[ \Delta_{ij}\sigma_{ij}^z
+ \left(V_{ij} \ket{g_i}\bra{e_j} + \text{H.C.} \right)\right].
\eeq
where $V_{ij} = F_{ij}e^{i\phi_{ij}}$ are the complex Rabi frequencies and
\beq
\label{eq:defDetuning}
\Delta_{ij} = E^{(e)}_j - E^{(g)}_i  - \omega_{ij}
\eeq
 is the detuning between the atomic transition energy $ E^{(e)}_j - E^{(g)}_i$ and the laser frequency $\omega_{ij}$.
$\gamma_{ij}D_{ij}(\rho)$ describes the decay of excited state $\ket{e_j}$ to ground state $\ket{g_i}$. According to the Lindblad equation~\cite{Lindblad1976,Gorini1976}:
\begin{equation}
\label{eq:DissipativeL}
D_{ij}(\rho) = -\frac{1}{2}\left\{\sigma^{\dagger}_{ij}\sigma_{ij}, \rho \right\}
+\sigma_{ij}\rho\sigma_{ij}^{\dagger}
\end{equation}
where $\left\{A,B\right\}$ is the anti-commutator of $A$ and $B$ and
$\sigma_{ij} = \ket{g_i}\bra{e_j}$.

We show in appendix~\ref{app:RWA} that in the RWA approximation this time-independent formulation is possible if and only if there is at most one laser coupling associated to each transition $g_i\leftrightarrow e_j$, as we have mentioned above, but also if to each such coupled transition, we can assign two real numbers  $\epsilon_i$, $\epsilon_j$  such that:
\beq
\label{eq:RWAConstraint}
\epsilon_i - \epsilon_j = \omega_{ij}.
\eeq
This condition cannot always be met for arbitrary frequencies and can impose a relation between the laser frequencies.

There are also important cases of interest where control fields are added within the ground or excited manifolds.
These can be resolved by block-diagonalizing them and redefining the  ground $\ket{g_i}$ or excited states $\ket{e_j}$ so that the problem is brought back to the desired form. Some of these schemes have been shown to be desirable for example in the acceleration of cooling of trapped ions \cite{Cerillo2011,Cerillo2018}.
\subsection{General conditions for the existence of dark states}
We say that a stationary state is dark when it involves only the ground-state manifold (there is no population in the excited states).
In general the steady-state has non-zero density on both the ground state $\{\ket{g_i}\}$ and excited state $\{\ket{e_j}\}$ manifolds, and only fulfills the conditions of dark states for specific values of the parameters.
The problem of tuning to dark states consists in finding the set of parameters, that is the laser intensities and frequencies, for which the steady state belongs to the ground state manifold $\{\ket{g_i}\}$.

It is convenient to  rewrite the Lindblad operator as a non-hermitian Hamiltonian part $L_{\tilde{H}}$, and a quantum jump operator $J$, as has been written often for example in the study of blinking or quantum trajectory theory~\cite{Plenio1998}:
\begin{equation}
L\rho= L_{\tilde{H}}\rho+ J(\rho)
\label{eq:LNH}
\end{equation}
where $L_{\tilde{H}}\rho =  -i(\tilde{H}\rho-\rho \tilde{H}^{\dagger})$ with $\tilde{H} = H - i\Gamma$ with
$\Gamma=\sum_{\left\{ij\right\}} \gamma_{ij} (\sigma_{ij}^{\dagger}\sigma_{ij})$ and
$J(\rho) = \sum_{\left\{ij\right\}} \gamma_{ij} \sigma_{ij} \rho \sigma_{ij}^{\dagger}$.

Our general method to obtain the conditions for dark states relies on the following theorem:
\begin{theorem}
\label{th1}
The $N$-level system whose evolution is governed by the Lindblad operator $L$ given by Eq.~\eqref{eq:Lindblad} has a dark state $\rho_{\text{d}}$ if and only if $L_{\tilde{H}}\rho_{\text{d}}=0$.
\end{theorem}
In other words, a dark state is an eigenstate of $L_{\tilde{H}}$ with a zero eigenvalue.
The proof is given on appendix~\ref{app:proofTh1}, we just give the heuristic of the proof here. It is based on two observations.
\begin{enumerate}
\item In general the eigenvalues of $L_{\tilde{H}}$ have a real part which is strictly negative, which is related to the fact that the eigenvalues of $\tilde{H}$ have a stricly negative imaginary part due to the total  decay rate $\sum_{i}\gamma_{ij}$ of excited states $\{\ket{e_j}\}$.
The only way to have a zero  eigenvalue is such that the corresponding eigenstate $\rho_d$ has no component on these decaying excited states $\{\ket{e_j}\}$.
\item If $L_{\tilde{H}}\rho_d = 0$ then
also $L\rho_d = 0$ as the jump  operator gives zero, $J(\rho_d) = 0$, on any state in the subspace spanned by $\{\ket{g_i}\}$.
\end{enumerate}
In that way, when $L_{\tilde{H}}\rho_{\text{d}} = 0$, we ensure that the corresponding eigenvector is a steady state of $L$ with no component in the excited states.
Hence, it is a dark state. The reverse is also true, all dark state $\rho_d$ fulfill $L_{\tilde{H}}\rho_d = 0$.

As dark states belong to  the subspace spanned by the ground states only, it is thus convenient to define the projection operator on this subspace
$P = \sum_{i=1}^{N_g}\ket{g_i}\bra{g_i}$ and its orthogonal complement $Q = \un-P = \sum_{j=1}^{N_e}\ket{e_j}\bra{e_j}$.
Where $\un$ is the identity operator in the Hilbert space $\mathcal{H}$ spanned by the $N$ states.
We also define super-projectors $\mathcal{P}$ and
$\mathcal{Q} = \un - \mathcal{P}$, where here $\un$ means the identity operator on the Liouville space of linear operators on $\mathcal{H}$.
These super-projectors acts as superoperator in the following way: $\mathcal{P}\rho = P\rho P$  and
 $\sQ \rho = P\rho Q +  Q\rho P + Q\rho Q$.

By inserting the identity $\sP + \sQ = \un$, between $L_{\tilde{H}}$ and $\rho$ in the equation $L_{\tilde{H}}\rho = 0$, and projecting  the resulting equation with the two super-projectors, we obtain the following two equations~:
\begin{align}
  \label{eq:projectLrho}
\sP \LH \sP \rho + \sP \LH \sQ \rho &=0 \nonumber \\
\sQ \LH \sP \rho + \sQ \LH \sQ \rho &=0.
\end{align}
As dark states belong entirely to the ground state manifold, we can enforce the condition
 $\sP \rho = \rho$ and thus $\sQ \rho=0$. In appendix~\eqref{eq:conditionH} we show that using these conditions, Eqs.~\eqref{eq:projectLrho} become:
\begin{align}
\label{eq:PHPrho}
[P H P, \rho] &= 0 \\
\label{eq:QHPrho}
QHP\rho &= 0.
\end{align}
From the first equation, Eq.~\eqref{eq:PHPrho}, we infer that there exists a common orthonormal basis of $P\mathcal{H}$ in which the matrix representation of $\rho$ and $PHP$ is diagonal.
But $PHP$ is diagonal in the $\{\ket{g_i}\}$ basis,
\begin{equation}
PHP = \sum_{i=1}^{N_g} \mathcal{E}_i\ket{g_i}\bra{g_i}
\end{equation}
where we have defined (see Eq.~\eqref{eq:HRWA} and Eq.~\eqref{eq:defDetuning})
\begin{equation}
  \label{eq:defEi}
\mathcal{E}_i = \sum_j\Delta_{ij}.
\end{equation}
Therefore, if the $PHP$ spectrum is non degenerate then the only solutions to Eq.~\eqref{eq:PHPrho} are matrices $\rho$ which are also diagonal in this basis.
But this is a trivial solution and in this case Eq.~\eqref{eq:QHPrho} implies that $V_{ij} = 0$.
Hence, non trivial solutions can arise if and only if $PHP$ has a degenerate spectrum.
This degeneracy condition translates to a constraint on the laser frequencies.
For instance, the requirement that two $PHP$ eigenvalues are equal,  $\mathcal{E}_i =\mathcal{E}_{i'}$, consists in a relation between the detunings
$\sum_j\left(\Delta_{ij} - \Delta_{i'j}\right) =0$ (by Eq.~\eqref{eq:defEi}), which translates into a relation between laser frequencies (see Eq.~\eqref{eq:defDetuning}).

Let us denote $P_s$, the orthogonal projectors on the eigen-subspace of dimendion $d_s$, associated to the $d_s - 1$ times degenerate eigenvalue $\mathcal{E}_s$. We have
$\sum_s P_s = P$, $P_sP_{s'} = \delta_{ss'}P_s$, and $\sum_s d_s = N_g$.
Then $PHP$ can be written as
\begin{equation}
PHP = \sum_s \mathcal{E}_s P_s,
\label{eq:degenerate-condition}
\end{equation}
and the dark states $\rho$ must have a block diagonal form~:
\begin{equation}
\label{eq:rhoBlockDiag}
\rho = \sum_{\{s;d_s>1\}} P_s\rho P_s =  \sum_{\{s;d_s>1\}} p_s\rho_s,
\end{equation}
where $\rho_s = P_s\rho P_s/\tr{P_s\rho P_s}$ is a $d_s\times d_s$ normalized density matrix, and
$p_s = \tr{P_s\rho P_s}$.

The second constraint given by Eq.~\eqref{eq:QHPrho} can now be written as~:
\begin{equation}
\label{eq:QHPscond}
\forall s; \quad  d_s>1 \quad  QHP_s\rho_s = 0.
\end{equation}
We deduce that $\rho_s$ can in principle be any positive, Hermitian, with trace one linear operator defined on $\ker{QHP_s}$, the kernel of $QHP_s \subset P_s\mathcal{H}$.

To conclude this section, we summarize: to have a dark state, $PHP$ must have a degenerate spectrum, this puts a constraint on the laser frequencies.
The dark state is a statistical superposition of states $\rho_s$, defined on  $\ker{QHP_s}$ where $P_s$ are the orthogonal projectors on the eigen-subspaces of $PHP$, corresponding to degenerate eigenvalues of $PHP$.
Depending on the dimension of $\ker{QHP_s}$ this last condition may or may not impose a condition
on the Rabi frequencies $V_{ij}$, this is the subject of the next section.
\subsection{Dimension, unicity and purity}
\label{sec:C}
In addition to the constraint on laser frequencies, the condition that $\rho_s$ must be defined on $\ker{QHP_s}$ can be satisfied if and only if the $\ker{QHP_s}$ is not empty,
$\dim{\ker{QHP_s}} \ge 1$.
By the rank-nullity theorem, $\dim{\ker{QHP_s}} + \rank{QHP_s} = \rank{P_s}=d_s$. Hence, a dark state can exist if and only if for at least one of the eigen-subspace $P_s{\mathcal{H}}$ of dimension $d_s$,
\beq
\label{eq:rankCond}
\rank{QHP_s} \le d_s -1.
\eeq
As $\rank{QHP_s}\le \rank{Q}=N_e$,  then two different cases are in order:
\begin{itemize}
\item {\bf case 1} $N_e\le d_s - 1$. In this case, the Eq.~\eqref{eq:rankCond} is always fulfilled, regardless of the values taken by the Rabi frequencies $V_{ij}$. The constraint on the laser frequencies $\omega_{ij}$, giving the degeneracy of $PHP$ and determining the dimension $d_s$ of the eigen-subspace is necessary and sufficient for the existence of a dark state.
We have supposed that the constraint imposed by the RWA approximation (Eq.~\eqref{eq:RWAConstraint}) have  already been fulfilled.
The $\Lambda$, the M~systems and the so-called ``fan"~\cite{Shore2014} or "multipod" systems~\cite{Ivanov2010}, belong to this case.
They will be discussed in more detail in the next section.
\item  {\bf case 2} $N_e \ge d_s$. In this case,  $\rank{QHP_s} \le N_e-1 <  N_e = \rank{Q}$.
Lowering the rank of $QHP_s$ to a lower value than $\rank{Q}$ can 	not be obtained for all value of the Rabi frequencies. In other words, the $V_{ij}$ must satisfy some relations such that the kernel $QHP_s$ be non-empty.
Therefore, in this case, the existence of a dark state requires that in addition to the laser frequencies $\omega_{ij}$,  the Rabi frequencies $V_{ij}$ must fulfill some constraints.
The specific case $N_g=N_e=2$ which belongs to this case, will be discussed with more details in the next section.
\end{itemize}

We see that in general, when  $\dim{\ker{QHP_s}}>1$, or when there is more than one  eigenvalues of $PHP$ which is degenerate, then multiple dark states may exist.
More specifically, the stationary dark state can be represented by any density operator $\rho$ defined on $\bigoplus_{\{s;d_s>1\}} \ker{QHP_s}$ (see Eq.~\eqref{eq:rhoBlockDiag} and Eq.~\eqref{eq:QHPscond}). That is,
 if $N_s=\dim{\ker{QHP_s}}$,  then a stationary dark state can be represented by any block-diagonal $M\times M$  density matrix, where $M = \sum_{\{s;d_s>1\}} N_s$, where the sum runs over all degenerate eigen-subspace $P_s\mathcal{H}$ of $H$, and each block is an $N_s\times N_s$ positive, Hermitian matrix.
Therefore, the stationary dark state which is reached asymptotically in time  will depend on the initial state. The dark state will be unique if and only if there is only one eigenvalue $\mathcal{E}_s$ which is degenerate, and $\dim{\ker{QHP_s}}=1$. We note that in this case the dark state is a pure state.
We conclude that, for dark states, unicity implies purity. Hence, if there is a mixed dark state then it is not unique. Indeed, suppose that the dark state is a mixed state
$\rho = p_1\ket{\psi_1}\bra{\psi_1} + p_2 \ket{\psi_2}\bra{\psi_2}$, where $p_1$ and $p_2$ are the eigenvalues of $\rho$ and $\ket{\psi_1}$, $\ket{\psi_2}$, are the two corresponding orthonormal eigenvectors. Because $\rho$ is a dark state, its two eigenstates must belong to $\ker{PHP}$. But then any linear combination or any statistical superposition of two states will fulfill the dark state condition Eq.~\eqref{eq:QHPrho}. Then, there is not a unique dark state.

A simple way to achieve a unique dark state is to tune the frequencies such that the dimension of the unique degenerate subspace $d_s$ fulfills the equality $N_e = d_s -1$.
As we are in the  case~1   ($N_e\le d_s- 1$), there is a dark state regardless of the values of the $V_{ij}$, but in addition, $\dim{\ker{QHP_s}} = d_s - \rank{PHQ} = d_s-N_e = 1$.
 $\rank{PHQ} = N_e = d_s-1$ because we suppose that each considered excited state $\ket{e_j}$
is coupled to at least one ground state $\ket{g_i}$ by a Rabi frequency $V_{ij}$.

This is why such M systems~\cite{Shore2014} have attracted attention as a generalisation of the very well known
 $\Lambda$ systems where $d_s = N_g = 2$ and $N_e = 1$. Specific example illustrating these general considerations will be given in the next section.
\section{Examples}

In this section we illustrate the preceeding discussion with four examples that illustrate the different cases from section \ref{sec:C}: a case with a unique stationary dark state, a case with a dark stationary subspace, and a case that is overspecified and whose dark state depends on the Rabi frequencies. The fourth example is the more complex system of the 11 hyperfine levels of $^{87}$Rb. 
For each case we will review the necessary condition to have a degenerate subspace and a non-zero kernel for $QHP$, and the resulting dark state subspaces.

\subsection{Example 1: unique stationary state for zigzag systems ($N_g=N_e+1$)}

\begin{figure}[h]
\includegraphics[width=0.5\textwidth]{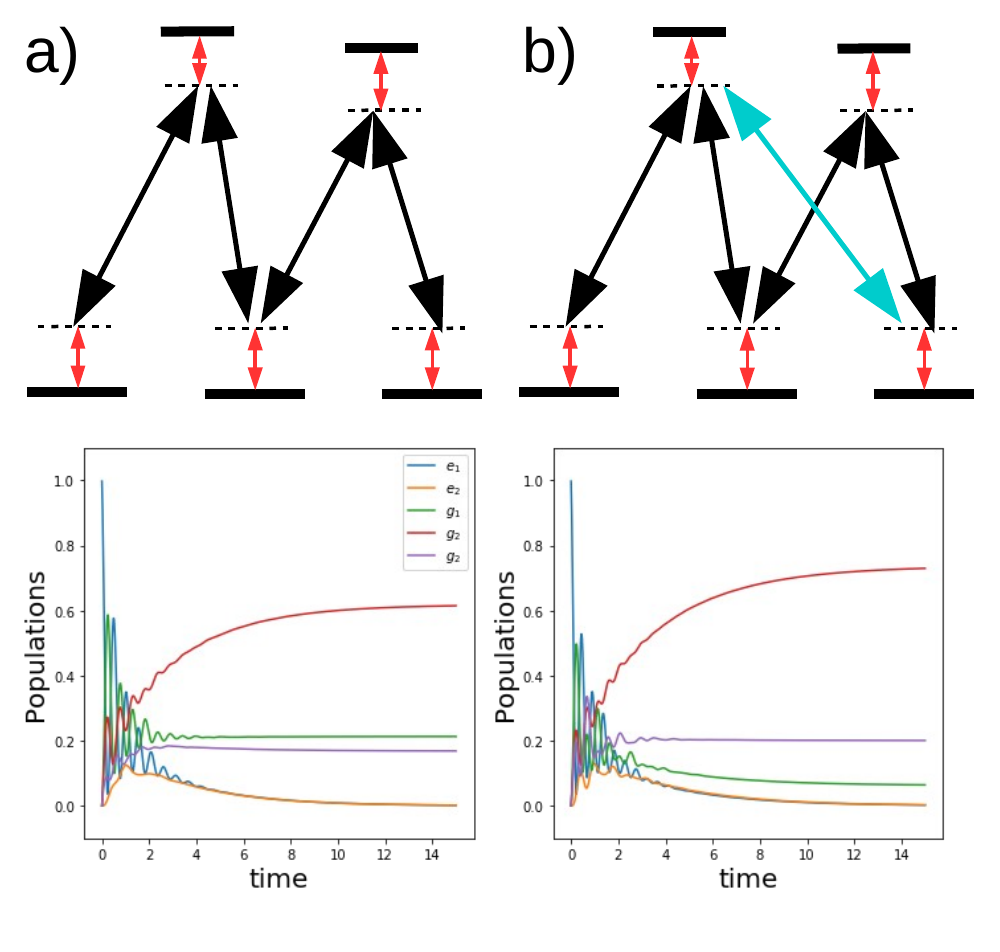}
	\caption{Populations of the $M$ system as a function of time in the a) Diagram of the $M$ system connectivity and probability amplitudes of the pure, dark, stationary state (red is positive probability amplitude, blue is negative). The dynamics of the system is plotted in the site basis and in the eigenstate basis. The system evolves towards a pure state occupying only the ground states. Parameters are $V_{21}=0.56, V_{22}=0.23, V_{32}=0.45, \gamma_{11}=0.04,\gamma_{12}=0.01,\gamma_{13}=0.09,\gamma_{21}=0.14,\gamma_{22}=0.02,\gamma_{23}=0.04$. b) Same as a) in the presence of the coupling $V_{31}=0.57$. In both cases the dark state is $\phi_5$ (all values in units of $V_{11}$).}
	\label{fig:M system with bias}
\end{figure}

We refer to zigzag or M systems as those where the connectivity given by the laser fields follows the pattern ground-(excited-ground)$_n$ (Figure \ref{fig:M system with bias}.a). We consider in particular the $M$ system with $N_g=3$ and $N_e=2$ which has been discussed elsewhere \cite{Gu2006}. It is a system where $N_g=N_e+1$ so that $\dim{\ker{QHP}}=1$ and $\dim{\ker{L}}=1$, provided that all detunings are equal. 
Figure \ref{fig:M system with bias}.a shows the evolution of the populations with initial conditions $\rho(t=0)=\ket{e_1}\bra{e_1}$ in the site basis $ \left\{ \ket{g_i};i=1,2,3 \right\},\left\{ \ket{e_j}; j=1,2 \right\}$ and in the eigenstate of $H$ basis $\left\{\ket{\phi_n}; n= 1,\cdots, 5\right\}$. As expected, the system evolves towards a pure dark state which here is $\ket{\phi_5}$.

Variations on the zigzag systems can be obtained by introducing additional coupling between ground and excited states. These modify the configuration of the stationary state over the ground state sites, but do not change its existence, uniqueness or purity. Because these additional couplings   create connectivity loops, the frequency of these additional laser fields cannot be independently chosen if we want to satisfy the RWA (Figure \ref{fig:M system with bias}.b and Appendix \ref{app:RWA}). In the particular case we are considering (see figure 2.b), the additional Rabi frequency $V_{31}$  must correspond to a laser frequency $\omega_{31}$ fulfilling: $\omega_{31}=\omega_{21}+\omega_{32}-\omega_{22}$.
The maximum number of  couplings  is $N_gN_e$, one for each couple $(g_i,e_j)$. Other control fields that fall beyond the scope of this article, are worth mentioning. For example, Cerillo et al. propose the addition of a control field between the ground states of a $\Lambda$ system as a means for accelerating the cooling rate \cite{Cerillo2011,Cerillo2018}. While the ground state can be prediagonalized and the magnitude of the ground-excited couplings redefined (thus leading to the starting point of this article), this has as a consequence to mix the restriction on detunings with restrictions on Rabi frequencies, resulting in dark states that depend on the intensity of the laser field as well. This point will be retaken in Example 3, where other examples of intensity dependent dark states are illustrated.

\subsection{Example 2: dark stationary subspace for fan systems ($N_g \geq 2,N_e=1$)}

Fan or multipod systems consist of $N_g$ ground states and a single excited state (the $\Lambda$ configuration is also an instance of a fan system although it has a unique stationary state).
In general the stationnary states of fan systems can be arranged in a number of degenerate subspaces, generated by pure states $\ket{\phi_n} = \sum_{i=1}^{d_s} c_{ni}\ket{g_i}$  where from Eq.~\eqref{eq:QHPrho}, the coefficients $c_{ni}$ must fulfill:
\begin{equation}
\sum_{i=1}^{d_s} V_{i1}c_{ni}=0
\label{eq:fan}
\end{equation}
where $V_{i1}$ are the Rabi frequencies of the laser coupling ground state $\ket{g_i}$ and the unique excited state $\ket{e_1}$.
Fan systems have stationary states of high multiplicity: for each subspace of dimension $d_s$ the kernel of the Liouvillian will have a dimension $(d_s-1)^2$ as well as $(d_s-1)^2$ conserved quantities (obtained as the left eigenvalues of the Liouvillian \cite{Albert2014, Kyoseva2006}.)

We specifically consider the case $N_g=4$. We begin with all four ground states forming a degenerate manifold and detune them one by one to assess its effect on the steady-state - whether it remains dark, pure and how its multiplicity changes. We choose equal couplings to the excited state for ease of visualization, and unequal relaxations back to the ground state ($V_{11}=V_{21}=V_{31}=V_{41}=1$, and $\gamma_{11}=0.1,\gamma_{12}=0.2,\gamma_{13}=0.3,\gamma_{14}=0.4$).
\newline

{\it a) Fully degenerate} ($d_1=4$). The fully degenerate
four-fold ground state case has $\dim{\ker{QHP}}=3$ and
$\dim{\ker{L}}=9$ (Fig. \ref{fig:example_4LS}.a).
A system initially in the  excited state will evolve towards a mixed $3\times 3$ density matrix (Tr$(\rho^2)=0.38$) determined by the conserved quantities that depend on the relaxation rates. Thus, although the condition to have the dark states depends exclusively on the Rabi frequencies and frequency detunings, the final reached state is a density matrix which depends also on the relaxation rates and on the initial state.

{\it b) One detuned ground state} ($d_1=3$). The system qualitatively similar to case a) has $\dim{\ker{QHP}}=2$ and $\dim{\ker{L}}=4$ (Fig. \ref{fig:example_4LS}.b) The evolution converges towards a $2\times 2$ density matrix corresponding to a mixed state (Tr$(\rho^2)=0.51$).

{\it c) Pairwise degenerate states} ($d_1=2$ and $d_2=2$).
Detuning a second ground state to the same value as the one of case b), results in pairwise degenerate levels (Fig. \ref{fig:example_4LS}.c). We must separately consider the degenerate subspaces, so we have $\dim{\ker{QHP_1}}=1$ and $\dim{\ker{QHP_2}}=1$, and dim(ker($L))=1^2+1^2=2$. The evolution converges to the
mixture of two pure states $p_1\ket{\phi_1}\bra{\phi_1} + p_2\ket{\phi_2}\bra{\phi_2}$, where  each $\ket{\phi_n}$ $(n=1,2)$ is the
stationnary state of a $\Lambda$ system. The weigths $p_n$ $(n=1,2)$ depend on the dissipation rates.
These are larger for the second manifold ($\ket{g_3},\ket{g_4}$) than for the first one ($\ket{g_1},\ket{g_2}$) and so the second manifold is more heavily populated.

{\it d) Minimum degenerate manifold} ($d_1=2$). The final detuning scheme keeps only a degenerate pair of levels (Fig. \ref{fig:example_4LS}.d). Then,  dim(ker($QHP_1))=1$ and dim(ker($L))=1$. Because the dark state is unique, it is also pure (see above). By properly choosing the detunings an experimentalist can localize in energy or space (if each energy level is spatially separated) via pumping.

We also note that because the excited states relax to the ground state, any dark state will accumulate all population and become the steady-state of the system. This is why in example d) the detuning of the other ground states does not render the state bright.

\begin{widetext}
\begin{center}
\begin{figure}[h]
\includegraphics[width=1.0\textwidth]{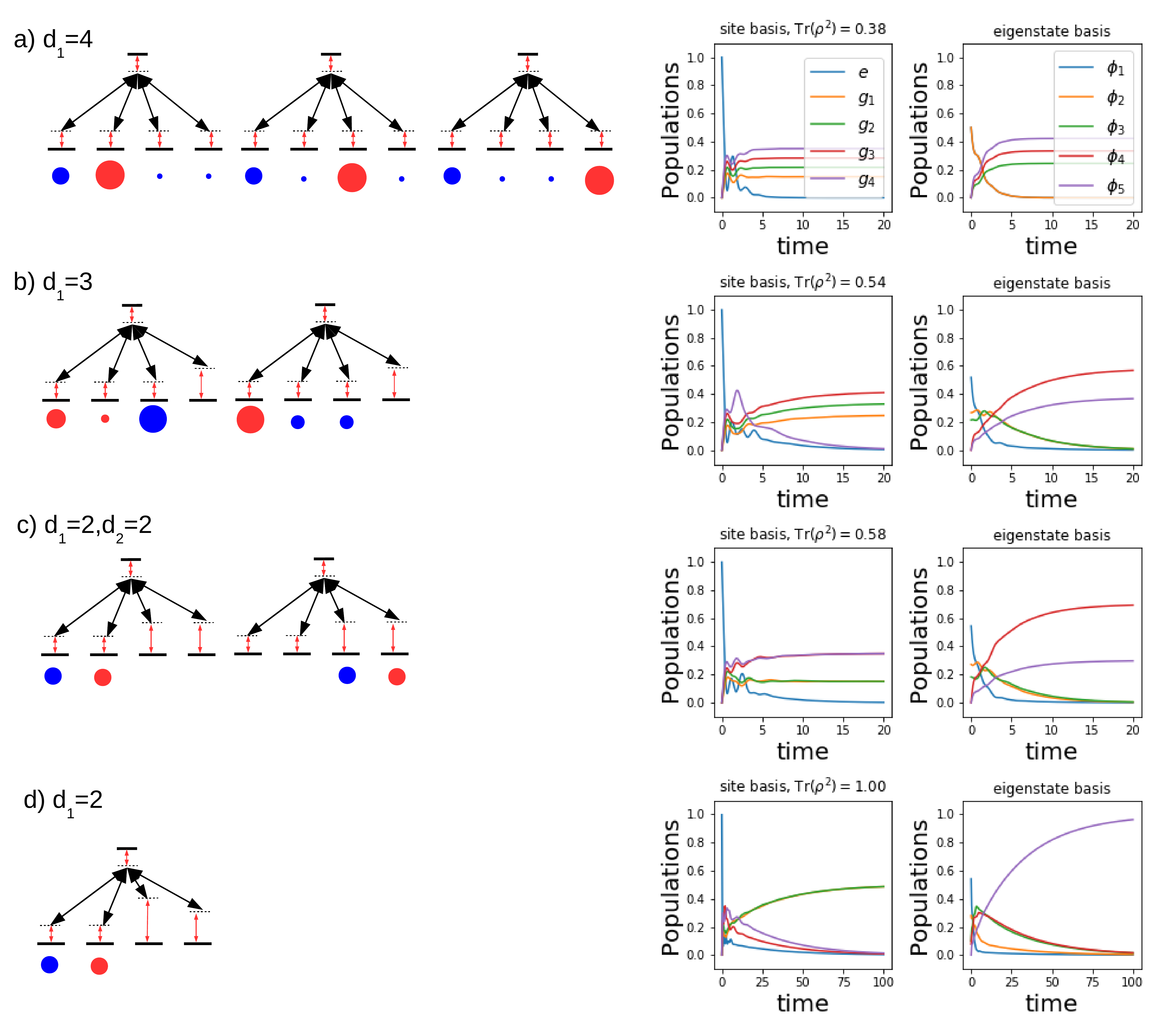}
	\caption{Tripod system with four ground state levels in different detuning configurations. Below the ystems are shown the wavefunctions that span the dark state subspace (blue means negative coefficient, red positive coefficient, size of the circle is proportional to the coefficient). On the left side are shown the evolution of the populations for an initially excited state. The purity of the steady-state is evaluated by calculating Tr$(\rho_{\text{SS}}^2)$. Parameters of the simulations are $V_{21}=V_{31}=V_{41}=1$, and $\gamma_{11}=0.1,\gamma_{12}=0.2,\gamma_{13}=0.3,\gamma_{14}=0.4$ (all values in units of $V_{11}$).}
	\label{fig:example_4LS}
\end{figure}
\end{center}
\end{widetext}

\subsection{Example 3: A Rabi frequency conditionned dark state: Pairs of two-level systems ($N_g \leq N_e$)}

We consider a system with two levels in the excited state and two levels in the ground state (Figure \ref{fig:example_4LS}). As
$d_s = 2$, we have  $\dim{\ker{QHP}}=2-\rank{QHP}$.  The generic case corresponds to $\rank{QHP} = \rank{Q}=2$
and in this case the kernel is empty - no dark state can exist in general, in contrast to the previous two examples. We will see that a dark state may exist but not for all values of the Rabi frequencies $V_{ij}$.
 Furthermore, the existence of a loop in the connectivity constrains the laser frequencies to fulfill the relation: $\omega_{21}=\omega_{11}+\omega_{22}-\omega_{12}$, such that the RWA results in a time independent Hamiltonian (see Appendix~A).

 The  Hamiltonian is (after a convenient referencing of the zero point energy):
\begin{equation}
H=\left[\begin{matrix}0 & 0 & V_{11} & V_{12}\\0 & \mathcal{E}_2 & V_{21} & V_{22}\\V_{11}^* & V_{21}^* & \mathcal{E}_3 & 0 \\ V_{12}^* & V_{22}^* & 0 & \mathcal{E}_4 \end{matrix}\right]
\end{equation}
The degeneracy condition implied by Eq.~\eqref{eq:PHPrho} requires that $\mathcal{E}_2=0$.
The constraint given by Eq.~\eqref{eq:QHPscond}
results in a relation between Rabi frequencies:
\begin{equation}
V_{11}V_{22}=V_{12}V_{21}.
\label{eq:constraint-fields-pairs}
\end{equation}

Figures~\ref{fig:pairs-2LS-dynamics}.a and \ref{fig:pairs-2LS-dynamics}.b  show the dynamics for a $\Lambda$ system and a $N_g=2,N_e=2$ system, respectively, to compare the effect of an additional excited state tuned to the dark state condition. Although the dynamical evolution differs slightly, the stationary state is identical. Increasing one of the Rabi frequencies gets the system out of the dark state condition onto a mixed stationary state where all four states are occupied, as shown in Figure~\ref{fig:pairs-2LS-dynamics}.c.

We can understand the setup and restrictions on the Rabi Frequencies by viewing the system as a pair of $\Lambda$ geometries on the same ground states. Because only one excited level is enough to fully specify the dark state condition (in general to fully specify a unique dark state in $N_g$ degenerate state one needs at least $N_g-1$ excited states), the second $\Lambda$ system must be adapted to  the ground state population specified by the first $\Lambda$ system. This can be achieved by tuning the Rabi frequencies.

\begin{figure}[h]
\includegraphics[width=0.5\textwidth]{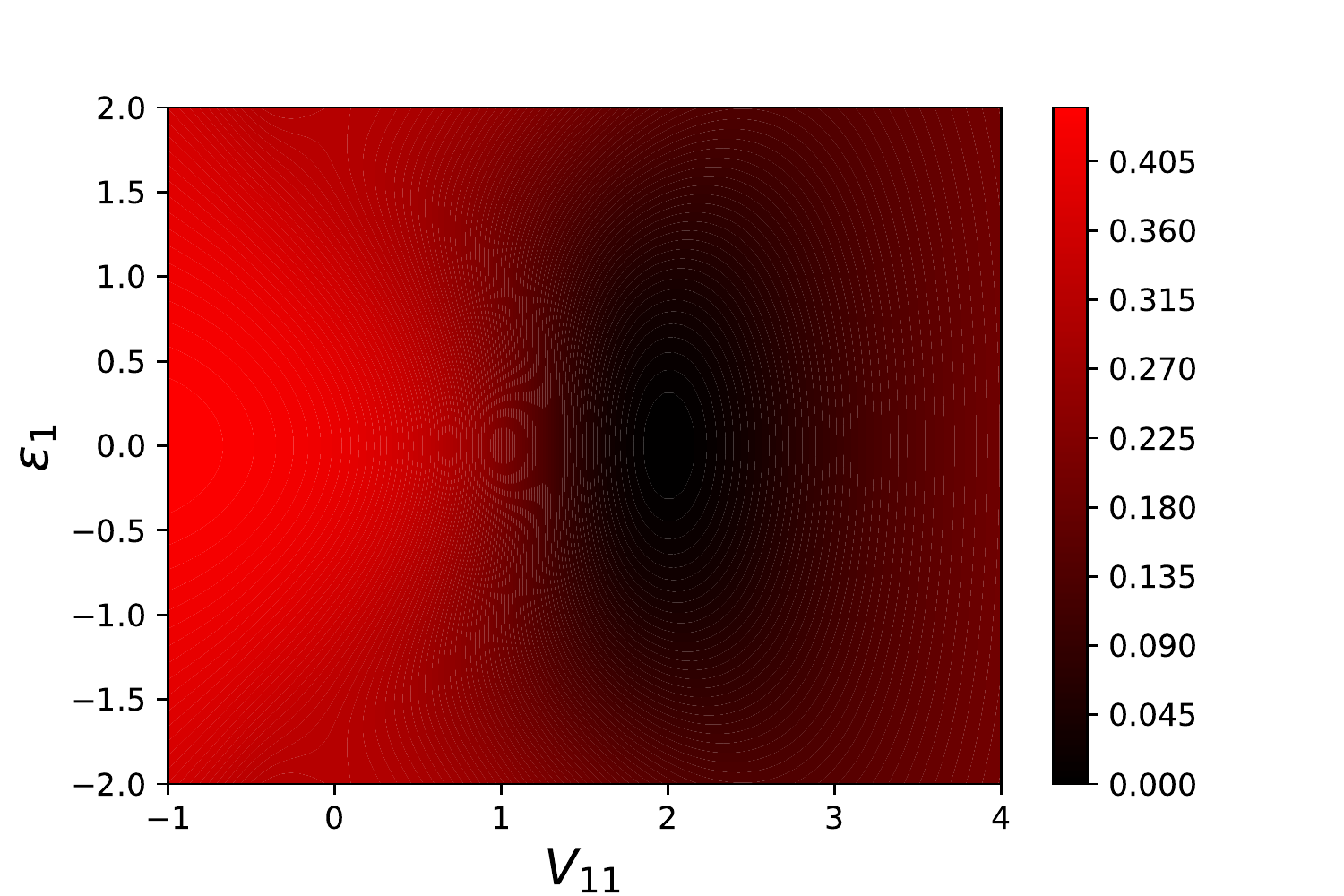}
	\caption{Excited state population as a function of detuning and field intensities for a pair of two-level systems. The parameters used are Parameters of the simulations are $V_{12}=2, V_{12}=1$, and $\gamma_{11}=0.2,\gamma_{12}=0.5,\gamma_{21}=0.44,\gamma_{22}=0.7$ (all values in units of $V_{22}$).}
	\label{fig:pairs-2LS}
\end{figure}

\begin{figure}[h]
\includegraphics[width=0.5\textwidth]{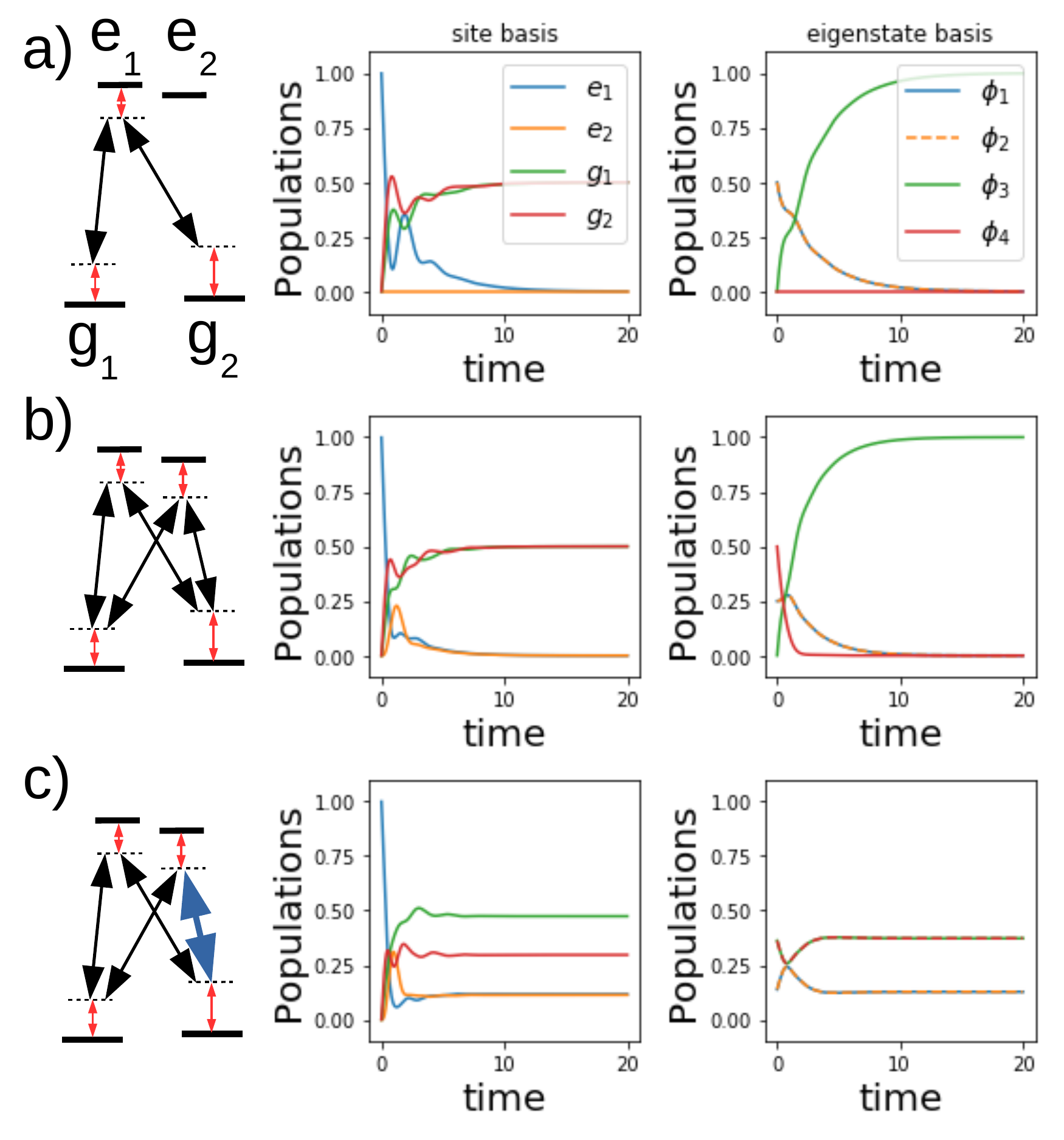}
	\caption{Excited state population for a pair of two-level systems in the case of a) Only couplings to one excited state ($V_{21}=1$) b) Coupling to both excited states in the dark state condition ($V_{21}=1$, $V_{12}=1$ and $V_{22}=1$) c) Coupling to both excited states not fulfilling the dark state condition ($V_{21}=1$, $V_{12}=1$ and $V_{22}=2$). The relaxation rates are $\gamma_{11}=0.2,\gamma_{12}=0.5,\gamma_{21}=0.44,\gamma_{22}=0.7$ (all values in units of $V_{11}$) and all detunings have been set equal.}
	\label{fig:pairs-2LS-dynamics}
\end{figure}

The map of excited state population vs. Rabi frequency and detuning (see Fig. \ref{fig:pairs-2LS}) reveals a dark state in frequency detuning as well as in Rabi frequency. The asymmetry in the map between positive and negative values of the Rabi frequency illustrates the phase sensitivity of the dark state.

Given a complex coupling $V_{ij}=F_{ij}e^{i\phi_{ij}}$, the requirement \eqref{eq:constraint-fields-pairs} separates into a constraint on the magnitudes of the fields $\abs{F_{11}}\abs{F_{22}}=\abs{F_{12}}\abs{F_{21}}$ and on the relative phases $\phi_{11}+\phi_{22}=\phi_{12}+\phi_{21}$. Such a dependence of the dark state on the Rabi frequencies could result in new metrology tools.

\subsection{Hyperfine splittings of $^{87}$Rb atoms}

\begin{figure}[b]
\includegraphics[width=0.5\textwidth]{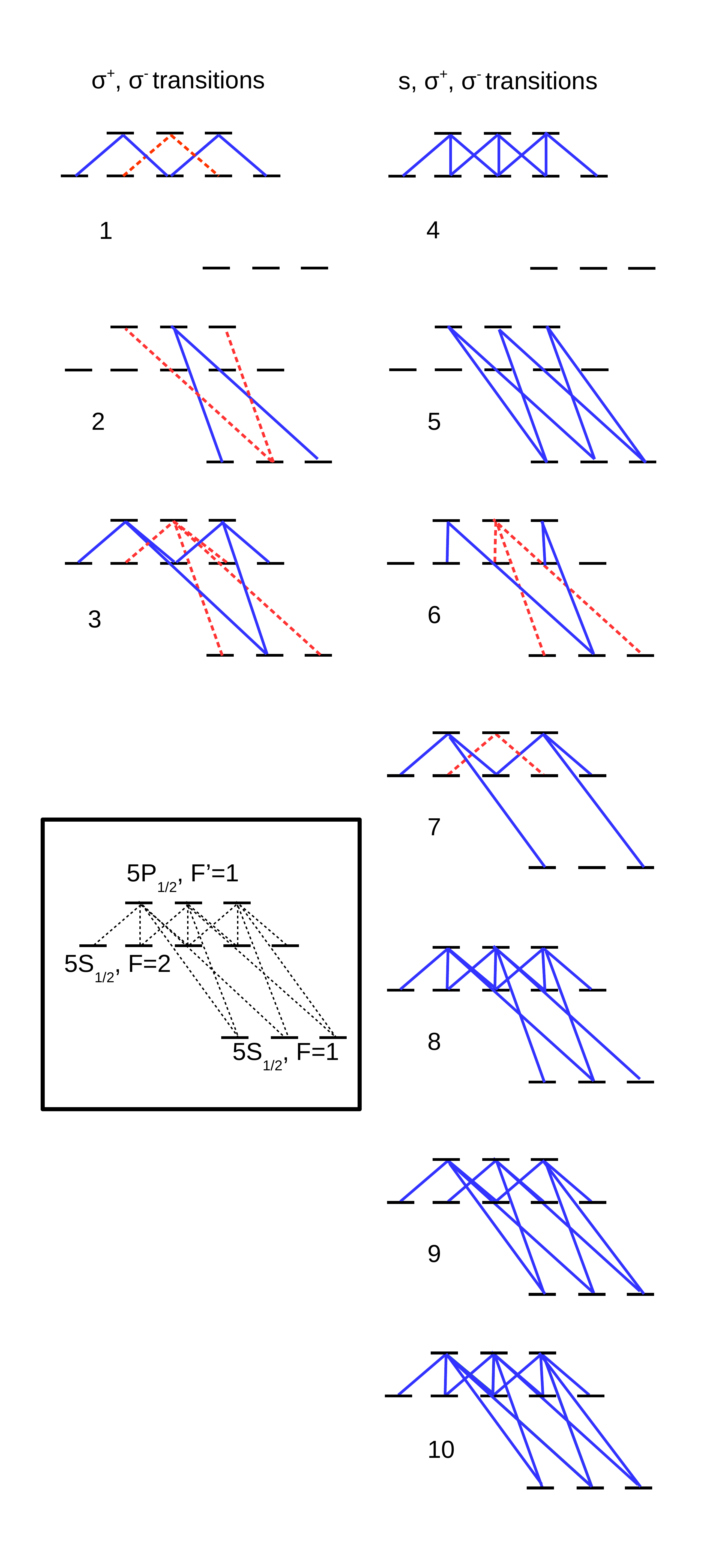}
	\caption{Possible connectivities for the hyperfine levels of $^{87}$Rb atoms. In the case of multiple manifold of dark states these are labeled by different colors (Blue (full) corresponds to $d_1$ and red (dashed) corresponds to $d_2$ of Table \ref{table:Rb87}). The case investigated in Ref. \cite{Ling1996} corresponds to case \textbf{3}.}
	\label{fig:Rb87}
\end{figure}

The characteristics of dark states outlined here allow a shortcut to establish their presence in more complex systems. As an example we take the study of CPT in multi-Zeeman-sublevel $^{87}$Rb atoms, as considered by  Ling et. al in Ref.~\cite{Ling1996}. In their publication,  Ling et. al  have investigated the possibility of obtaining CPT using only two co-propagating linearly polarized lasers addressing the two transitions that connect the higher energy level $P_{1/2} F'=1$ to the two lower levels $S_{1/2}, F=2$ and $S_{1/2}, F=1$, respectively. The main point is that each of these 3 levels has a hyperfine sublevel structure. Indeed, the $S_{1/2}, F=2$ is composed by 5 degenerated sublevels ($m_F = -2,-1,\cdots, 2$), and both $P_{1/2} F'=1$  and $S_{1/2}, F=1$ are composed by 3  degenerated sublevels each ($m_F = -1,0, 1$). Taking the quantization axis as the propagation direction, results in $\sigma^+$, $\sigma^-$ transitions with equal strength, as both lasers are linearly polarized~\cite{Ling1996}. The energy levels as well as all possible transitions ($\sigma^+$, $\sigma^-$ and $s$) are shown in Figure \ref{fig:Rb87} (boxed scheme). 

We will expand the scope of possible transitions beyond those considered by Lin et al. by introducing additional lasers that can induce $\Delta m_F=0$ and $\Delta m_F=\pm 1$ transitions. The selected schemes are shown in Figure \ref{fig:Rb87}. 
We apply to these cases the criteria developed in this article to obtain the number, dimension and purity of dark states without the need for expensive numerical simulations. 
 
As a first step we identify the number of degenerate subspaces that can independently sustain dark states (see Eq. \eqref{eq:degenerate-condition}). These degenerate subspaces can occur on account of different detunings (see e.g. Figure \ref{fig:example_4LS}), or because the connectivity results in two decoupled (ground-excited) systems such as in scheme 1, 3, 6 and 7 of Figure~\ref{fig:Rb87}.
For simplicity we assume in our analysis that all detunings are equal. The next step is to calculate the kernel of $QHP_s$ (see Eq. \eqref{eq:QHPscond}). 
The dimension of the resulting dark state manifold is $\text{dim}[\text{ker}(QHP_s)]=d_s-\text{dim}[\text{rank}(QHP_s)]$. For the connectivities considered, if $N_g>N_e+1$, there will always be a dark state for any field intensity, provided that the detuning condition is met 
(we note that $N_g$ and $N_e$ are only defined with respect to coupled states so that, for instance, $N_g = 5$ in Scheme~1 of figure~\ref{fig:Rb87} and the sublevels of the $F=1$ manifold does not contribute to $N_g$) These are the cases of all schemes except 5. In this case, we need that $\text{dim}[\text{rank}(QHP_s)]<N_e$, which can be obtained in the case by adjusting the intensities. The conditions are $- V_{11} V_{23} V_{32} = V_{12} V_{21} V_{33}$, where we have used the nomenclature of this paper where the first index labels a ground state and the second an excited state. In the specific case of the hyperfine transitions of $^{87}$Rb this transforms into $- V_{s} V_{\sigma^+} V_{\sigma^-} = V_{\sigma^+} V_{\sigma^-} V_{s}$, which has no solution given that different $\Delta m_F=\pm 1$ transitions cannot be independently controlled (this might be different for other physical systems of three ground-excited pairs). 
We gather in table~\ref{table:Rb87} our results. The existence of a dark state, its dimension and purity are indicated, as well as the levels spanned by the state. From this table more involved calculations such as have been carried out for $\Lambda$ systems can more seamlessly be extended \cite{Renzoni1999,Renzoni1999b,Failache2007,Choi2014}.

Having such a summary can be of further use since one can imagine preparing a pure dark state in the red (dashed) subsystem of scheme 1 (population in levels $F=2, m_F=-1,+1$) by first preparing the dark state of scheme 2 ($F=1, m_F=-1,+1$), and adiabatically turning on the $F'=1$ to $F=2$ transition to transfer populations adiabatically to the $F=1, M=-1,+1$ levels, and turning off the laser for the $F=1$ to $F'=1$ transition to result in a pure state of scheme 1 ($F=2, M=-1,+1$), without any population in the $F=2, M=-2,0,+2$ dark state. Such a state is not obvious to prepare, but becomes more transparent from our analysis. 
We remark that for several of these cases non-idealities play different roles. Collisional relaxation among the ground state manifold induces a small population in the excited states. We also remark that in the cases where certain states are uncoupled by light but coupled by dissipative pathways (1,4,5,6,7), some of the population might be lost to them (i.e., sink states).

\begin{table*}
\resizebox{\textwidth}{!}{\begin{tabular}{ | c| c| c| c| c| c| c| l c c c | l c c c|}
\hline	&		&		&		&		&		&		&	$d_1$	&		&		&		&	$d_2$	&		&		&		\\
	&	s	&	$\sigma^+$	&	$\sigma^-$	&	s	&	$\sigma^+$	&	$\sigma^-$	&	states	&	dim	&	purity	&	cycles	&	states	&	dim	&	purity	&	cycles			\\	\hline
1	&		&	x	&	x	&		&		&		&	F=2, M=-2,0,+2	&	1	&	yes	&	0	&	F=1, M=-1,+1	&	1	&	yes	&	0	\\
2	&		&		&		&		&	x	&	x	&	F=1, M=-1,+1	&	1	&	yes	&	0	&		&		&		&				\\
3	&		&	x	&	x	&		&	x	&	x	&	F=2, M=-2,0,+2 / F=1, M=0	&	2	&	no	&	1	&	F=2, M=-1,+1 / F=1, M=-1, +1	&	3	&	no	&	0	\\
4	&	x	&	x	&	x	&		&		&		&	F=2, M=-2,-1,0,+1,+2	&	2	&	no	&	2	&		&		&		&			\\
5	&		&		&		&	x	&	x	&	x	&	F=1, M=-1,0,+1	&	1	&	yes	&	1	&		&		&		&		\\
6	&	x	&		&		&		&	x	&	x	&	F=2, M=-1,+1 / F=1, M=0	&	1	&	yes	&	0	&	F=2, M=0 / F=1, M=-1,+1	&	2	&	no	&	0	\\
7	&		&	x	&	x	&	x	&		&		&	F=2, M=-2,0,+2 / F=1, M=-1, +1	&	3	&	no	&	0	&	F=1, M=-1,+1	&	1	&	yes	&	0	\\
8	&	x	&	x	&	x	&		&	x	&	x	&	F=2, M=-2,-1,0,+1,+2 / F=1, M=-1,0,+1	&	5	&	no	&	3	&		&		&		&		\\
9	&		&	x	&	x	&	x	&	x	&	x	&	F=2, M=-2,-1,0,+1,+2 / F=1, M=-1,0,+1	&	5	&	no	&	2	&		&		&		&		\\
10	&	x	&	x	&	x	&	x	&	x	&	x	&	F=2, M=-2,-1,0,+1,+2 / F=1, M=-1,0,+1	&	5	&	no	&	5	&		&		&		&		\\	\hline
\end{tabular}}
\caption{Characteristics of the dark states corresponding to the systems illustrated in Figure \ref{fig:Rb87}.
The first 3 columns indicate the polarization of the laser inducing the $F=2\leftrightarrow F'=1$ transition and the next 3 columns are for the laser polarization coupling the states corresponding to the  $F=1\leftrightarrow F'=1$ transition. In some case 2 manifolds of dark states can take place: columns $d_1$ and $d_2$. In each $d_1$ and $d_2$ column, the atomic states ivolved, the dimension, purity and the number of cycles of the dark states are indicated. Scheme 5 is marked as a dark state, although the conditions imposed on the intensities cannot be fulfilled with the radiative transitions' degrees of freedom considered here (see text for details).}
\label{table:Rb87}
\end{table*}

\section{Conclusions}

We have presented an overview of dark states conditions on dissipative systems classified as a function of the number of ground and excited states. The condition can be reduced to a condition on the Hamiltonian part of the evolution. The number of excited and ground states naturally separates the systems into a case where CPT depends only on the detunings, and one where CPT appears only once conditions on the detunings and Rabi frequencies are met. When the kernel has multiplicities higher than one, the stationary states are mixed and the term coherent population trapping becomes less apt. Conserved quantities determine the final state and these depend on the dissipative rates of the system.
\newline

\section{Acknowledgements}

DFS acknowledges a Marie-Sklodowska-Curie Fellowship.

\appendix
\section{Rotating wave approximation}
\label{app:RWA}
We use throughout the article the rotating wave approximation (RWA) to turn the time-dependent Hamiltonian into a time-independent Hamiltonian.
For $N$-level systems with arbitrary connectivity this may impose important constraints on the wavelengths or detunings that can be used.

Let start from the original time-dependent Schrodinger equation for the unitary evolution operator generated by the time dependent Hamiltonian $H(t)$,
$i\dot{U}(t) = H(t) U(t)$,
where $H(t)$ can be  written as:
\beq
H(t) = \sum_{i=1}^NE_i \ket{i}\bra{i} +
\sum_{i>j=1}^N F_{ij}(t)\left(\ket{i}\bra{j} + \ket{j}\bra{i}\right)
\eeq
and $F_{ij}(t) = 2F_{ij} \cos(\omega_{ij}t +\phi_{ij})$, where the $F_{ij}$, $\omega_{ij}$ and and $\phi_{ij}$ are time-independent constants.

We look for a diagonal operator
$H_0 = \sum_{i=1}^{N} \epsilon_i \ket{i}\bra{i}$,
such that after
applying the unitary operator $U_0 = e^{-iH_0 t}$, the original time dependent Schr\"{o}dinger equation becomes time-independent within a good approximation.

Specifically, let write $U(t) = U_0(t)U_I(t)$, then $U_I(t)$ fulfills the Schr\"{o}dinger
$i\dot{U}_I= H_I(t) U_I(t)$, where $H_I = \sum_{i=1}^N \left(E_i-\epsilon_i\right) \ket{i}\bra{i} + V(t)$ where $V(t)$ is given by:
\beq
V(t) = \sum_{i>j=1}^N F_{ij}(t)\left(e^{i\left(\epsilon_i-\epsilon_j\right)t}\ket{i}\bra{j} + \text{H.C}\right),
\eeq
where H.C means the hermitian conjugate of the preceding term.

Using the explicit expression of $F_{ij}(t)$ as  $F_{ij}(t) = \left(V_{ij}e^{i\omega_{ij}t} + V_{ij}^*e^{-i\omega_{ij}t} \right)$, where $V_{ij} = F_{ij}e^{i\phi_{ij}}$, we obtain:
\begin{align}
\label{eq:interacPicture}
V(t) &= \sum_{i>j=1}^N
 \left( V_{ij}e^{i\left(\epsilon_i-\epsilon_j - \omega_{ij} \right)t} \ket{i}\bra{j} + \text{H.C} \right) \nonumber \\
& +
 \left( V_{ij} e^{i\left(\epsilon_i-\epsilon_j + \omega_{ij} \right)t} \ket{i}\bra{j} + \text{H.C}
\right)
\end{align}
The RWA consists in choosing a set of energies $\epsilon_i$ such that
\beq
\label{eq:RWAcondition}
\epsilon_i-\epsilon_j = \omega_{ij}
\eeq
and such that the terms oscillating at the frequencies
$\epsilon_i-\epsilon_j\ + \omega_{ij} = 2 \omega_{ij}$, corresponding to the third line of Eq.~\eqref{eq:interacPicture},  can be safely neglected.

Within this aproximation, $U_I(t)$ fullfils a time-independent Schr\"{o}dinger equation,
$\dot{U}_I \simeq H_IU_I(t)$, where $H_I$ is time independent and given by,
\beq
H_I = \sum_{i=1}^N (E_i-\epsilon_i) \ket{i}\bra{i}
+ \sum_{i>j=1}^N \left(V_{ij} \ket{i}\bra{j} + \text{H.C} \right).
\eeq
It is convenient to introduce the operators
$\sigma_{ij}^z = \ket{i}\bra{i} - \ket{j}\bra{j}$ and take advantage of the condition Eq.~\eqref{eq:RWAcondition},
 in order to rewrite $H_I$ as~:
\beq
\label{eq:appHRWA}
H_I = \sum_{i>j=1}^N \left[ \Delta_{ij}\sigma_{ij}^z
+ \left(V_{ij} \ket{i}\bra{j} + \text{H.C} \right)\right].
\eeq
where we have defined the detunings
$\Delta_{ij} = E_i-E_j -\omega_{ij}$ and where we have ignored a term proportional to the identity which corresponds to an arbitrary energy origin.

The unitary transform $U_0$ does not affect the dissipative part of the Lindblad operator (Eq.\eqref{eq:Lindblad})
 Indeed,
$\rho_I = U_0^\dagger \rho U_0$ fulfill the Lindblad equation
$\dot{\rho}_I = -i[H_I,\rho_I] + \sum_{\{ij\}} \gamma_{ij} D_{ij}(\rho)$ as
$U_0^\dagger D_{ij}(\rho) U_0 = D_{ij}(\rho)$ (see Eq.~\eqref{eq:DissipativeL}).
Therefore within the RWA approximation, the Lindblad operator is  time independent.

We note that this is possible only if we can find a solution of
Eq.~\eqref{eq:RWAcondition}, giving the $N$ unknowns $\epsilon_i$, as a function of the laser frequencies $\omega_{ij}$.
We infer that in general the number of equation must be less than $N$. But this constraint is not sufficient. Even when the number of equations is less than $N$, additional constraints on the laser frequencies $\omega_{ij}$ may be imposed to obtain a solution of Eq.~\eqref{eq:RWAcondition}. This is always the case when there is a closed cycle in the set of couple $(i,j)$.
For example, consider a N-level system with $N\ge 3$, where the transitions $(i,j) = (1,2)$, $(2,3)$ and $(1,3)$ are driven by 3 different laser fields.
Then, summing Eq.~\eqref{eq:RWAcondition}, over $i$ and $j$, with $i>j$, gives $\omega_{12} +\omega_{23} + \omega_{13} = 0$.
\section{Proof of theorem~\ref{th1}}
\label{app:proofTh1}
\begin{proof}
If $\det[L_{\tilde{H}}]=0$, then $L_{\tilde{H}}$ has a right eigenvector $\rho_d$ which fulfills $\tilde{H}\rho_d-\rho_d\tilde{H}^\dagger=0$. $\rho_d$ is a steady state of the dynamics generated by $L_{\tilde{H}}$.
We separate explicitly the Hermitian and non-Hermitian part of $\tilde{H}$ and this condition becomes:
\begin{equation}
\label{eq:LHequal0}
[H,\rho_d]-i\{\Gamma,\rho_d\}=0
\end{equation}
where $\Gamma$ is given by
$\Gamma=\sum_{\left\{ij\right\}} \gamma_{ij} (\sigma_{ij}^{\dagger}\sigma_{ij})$
with $\sigma_{ij} = \ket{g_i}\bra{e_j}$ or
\beq
\label{eq:Gammaj}
\Gamma = \sum_{j=1}^{N_e} \Gamma_{j}\ket{e_j}\bra{e_j}
\eeq
with $\Gamma_{j} = \sum_{i=1}^{M} \gamma_{ij}>0$ is the total decay rate of the excited state $\ket{j}$.
Hence, $\Gamma$
is a completely positive diagonal matrix and the ground states manifold $\{\ket{g_i}\}$ is a subspace of its kernel, $\Gamma\ket{g_i} = 0; \forall i=1,2,\cdots,M$.

Taking the trace of Eq.~\eqref{eq:LHequal0}, gives us $\tr{\Gamma\rho_d} = 0$, as the trace of the commutator gives zeros. Using Eq.~\eqref{eq:Gammaj}, this last condition can be written as~:
\beq
\sum_{j=1}^{N_e} \Gamma_j\bra{e_j}\rho_d\ket{e_j}=0.
\eeq
But each term of this sum is positive, therefore each term must be zero, thus
$\bra{e_j}\rho_d\ket{e_j} = 0;\forall j=1,2,\cdots,N_e$.
We conclude that $\rho_d$ has no population in the excited state manifold, therefore it can neither have coherence involving excited states,
$\bra{e_j}\rho_d\ket{e_{j'}} = \bra{e_j}\rho_d\ket{g_{i}} = 0$.
We conclude that $\rho_d$ lies in the ground state manifold $\{\ket{g_i}\}$.

Now we consider the original Lindblad operator $L$ which includes the quantum jump operator $\sum_{ij} \gamma_{ij} \sigma_{ij} \rho_d \sigma_{ij}^{\dagger}$, written explicitely as~:
\beq
J(\rho_d) = \sum_{ij}\ket{g_i}\bra{e_j}\rho_d\ket{e_j}\bra{g_i} = 0
\eeq
where we have used $\bra{e_j}\rho_d\ket{e_j}=0$. Therefore $L\rho_d = 0$.

Finally $\rho_d$ is a steady state of $L$ with no component in the excited state manifold, hence it is a dark state.
\end{proof}
We have thus proved that if $L_{\tilde{H}}\rho =0$ then $\rho$ is a dark state. The converse, is obviously true.

\section{From super-projector to projector}
\label{eq:conditionH}
Starting from equations
\begin{align}
\sP \LH \sP \rho + \sP \LH \sQ \rho &=0 \nonumber \\
\sQ \LH \sP \rho + \sQ \LH \sQ \rho &=0.
\end{align}
we would like to prove their equivalence with the following equations:
\begin{align}
[P H P, \rho] &= 0 \\
QHP\rho &= 0.
\end{align}

We first enforce $\rho = \mathcal{P}\rho$ and $\mathcal{Q}\rho = 0$. Then:
\begin{align}
\sP \LH \sP \rho &=0 \nonumber \\
\label{eqapp:conditions}
\sQ \LH \sP \rho &=0.
\end{align}

It is convenient to introduce the column form $\rho_s$ of $\rho$, which convert an
$n\times n$ matrix $\rho$, to a $N^2$ column vector $\rho_s$.
In this transformation, the operation $A\rho B^{\dagger}$ is mapped to $\bar{B} \otimes A \rho_s$,
where  $\bar{B}$ denote the  conjugate of $B$~\cite{Havel2003}.

The effect of this mapping on the super-projector is as follows:
\begin{align}
  \label{eq:PQmap}
\mathcal{P}\rho & \rightarrow P \otimes P \rho_s \nonumber \\
\mathcal{Q}\rho & \rightarrow  \left(P\otimes Q + Q\otimes P + Q\otimes Q\right) \rho_s.
\end{align}
The superoperator $L_{\tilde{H}}$ is mapped as:
\beq
\label{eq:Lmap}
L_{\tilde{H}}\rho \rightarrow -i\left(\un \otimes \tilde{H} -
\tilde{H}^{t} \otimes \un \right)\rho_s
\eeq
where $\tilde{H}^{t}$ is the transposed of $\tilde{H}$.
Using Eqs.~\eqref{eq:PQmap} and Eq.~\eqref{eq:Lmap}, we can map Eqs.~\eqref{eqapp:conditions} as:
\begin{align}
  \label{eq:mapPHP}
\left(P \otimes PHP - PHP \otimes P\right)\rho_s = 0 \\
\label{eq:mapQHP}
\left(P \otimes QHP - Q\tilde{H}^t P \otimes P \right)\rho_s = 0,
\end{align}
where we have used that $P\tilde{H}P = P\tilde{H}^tP =PHP$ and $Q\tilde{H}P = QHP$.
By reversing the mappping, the first equation (Eq.~\eqref{eq:mapPHP}) gives
\beq
[PHP,P\rho P] = 0
\eeq
which is Eq.~\eqref{eq:PHPrho} of the main text.

For the second equation (Eq.~\eqref{eq:mapQHP}) we use that the dark state fulfills $\rho = P\rho P$. Therefore, the second term $Q\tilde{H}^t P \otimes P \rho_s = 0$. Indeed, by reversing the mapping, it can be written as $P \rho P Q\tilde{H}^{\dagger}P = 0$.
Hence, Eq.~\eqref{eq:mapQHP} is equivalent to $P \otimes QHP \rho_s = 0$, which, by reversing the mapping,  gives :
\beq
QHP P\rho P = 0,
\eeq
which is Eq.~\eqref{eq:QHPrho} of the main text.

\bibliography{Fano,dark-states}

\end{document}